\documentclass{article}
\usepackage{iftex}
\ifPDFTeX
  \usepackage[utf8]{inputenc}
\fi
\usepackage[T1]{fontenc}
\usepackage{genai4health_2026}
\usepackage{hyperref}
\usepackage{url}
\usepackage{booktabs}
\usepackage{amsmath,amssymb}
\usepackage{graphicx}
\hypersetup{hidelinks}

\title{Co-Defining Transparency for Clinical AI: A Replicable Participatory Methodology}

\author{%
  Mina Bjelogrlic$^{1,2, *,\dagger}$ \quad
  Inês Espinhal Amaro$^{1,*}$ \quad
  Hugues Turbé$^{1,2}$ \quad
  Julie Guebey$^{1}$ \\
  \bfseries Christian Lovis$^{1,2}$ \quad
  Laëtitia Gosetto$^{1,2}$ \\[0.6em]
  $^{1}$ Division of Medical Information Sciences, Geneva University Hospitals, Switzerland\\
  $^{2}$ Department of Radiology and Medical Informatics, University of Geneva, Switzerland\\[0.4em]
  {\small $^{*}$Equal contribution \quad
  $^{\dagger}$Corresponding author: \texttt{mina.bjelogrlic@hug.ch}}
}

\begin{document}
\maketitle

\begin{abstract}
Transparency is the most invoked safeguard for clinical AI, yet frameworks (e.g.\ the EU AI Act, FUTURE-AI) are specified top-down by regulators and technical experts, not bottom-up by the frontline users whose trust and adoption they should calibrate. This disconnect exacerbates the central deployment failure of clinical AI, \emph{miscalibrated reliance and adoption} (such as automation bias, deskilling, or disuse), which technical explanation alone does not resolve. We argue that ``how much transparency, of what kind, and for whom'' is an empirical, context-dependent question, and present a replicable participatory methodology to answer it, instantiated at a large European university hospital. A cumulative chain of six workshops engages an inter-professional panel (such as clinicians, patient and caregiver partners, and students) in agnostic, open-ended elicitation: 275 stakeholder verbatims and external benchmarks (eCH-0272, EU AI Act, Stanford FMTI) yield 20 indicators (18 prioritised and 2 recovered at later stage), rated across five clinical scenarios of graduated intervention before and after technical briefings, discussed with legal experts, and translated into concrete implementation features. As an existence proof, the use case yields decision-useful signals: the human review of automated decisions is the single most critical transparency indicator (84\% of participants), and required transparency for adoption rises monotonically with the perceived risk of intervention (AI informational tools to AI treatment tools), corroborating World Health Organisation (WHO) risk-proportionality. In the local setting, a short explainability technical briefing did not substantially shift participants' transparency requirements. Of 151 evaluated AI transparency solutions, 79 were accepted as-is, covering (at least partially) 17 out of the 20 selected indicators. We position the six-workshop protocol (not its single-site outcomes) as the transferable artefact and release all instruments to support replication.
\end{abstract}

\section{Introduction}
Generative and predictive AI is being embedded across the clinical pathway, from patient-facing chatbots and ambient scribes to triage, diagnosis and treatment recommendation. Across ethics guidelines, regulation and governance strategies, \emph{transparency} is the single most cited safeguard \citep{franzke2022,kiseleva2022}. Yet transparency is not one thing: depending on context it denotes data provenance, model interpretability, traceability, uncertainty communication, or disclosure of limitations, each with distinct technical and organisational costs. International references such as the EU AI Act \citep{euaiact2024}, the Council of Europe Convention \citep{coe2024}, the FUTURE-AI guideline \citep{futureai2025} and the Foundation Model Transparency Index \citep{fmti2024} give valuable normative structure, but they are expert-driven and top-down: they prescribe what transparency \emph{should} be rather than establish what it \emph{must accomplish} for the people who act on the system's output.
 
The purpose transparency must accomplish, we argue, is to calibrate reliance, trust and adoption. The dominant deployment problem for clinical AI is not raw accuracy but whether the human relies on the system \emph{when and only when} it is right \citep{leesee2004,schemmer2023}. Both failure modes are documented in medicine: over-reliance through automation bias and anchoring \citep{khera2023,jabbour2023}, longer-term deskilling under routine AI exposure \citep{budzyn2025}, and (less discussed but empirically more prevalent in some settings) under-reliance/disuse, where clinicians discard correct advice \citep{calibrating2026,kuper2025}. Crucially, adding explanations does not reliably repair this: explanations can fail to reduce over-reliance and sometimes increase it \citep{bansal2021,bucinca2021}, help only when they lower the cost of verification \citep{vasconcelos2023}, and do not neutralise reliance on \emph{biased} models even for AI-literate physicians \citep{jabbour2023,qazi2026}. AI Transparency is therefore one important element impacting on the clinical stake, the user, and the healthcare workflow, which should be carefully designed and evaluated. The co-designed AI transparency solutions should later be evaluated in prospective randomised controlled trials to assess, on clinically relevant indicators such as Patient Reported Outcomes (PROMs) or TTF (time-to-treatment-failure), whether this additional transparency is beneficial to the healthcare system.
 
Transparency perception is contingent; it cannot be settled by a fixed checklist; we propose to elicit it from stakeholders against concrete uses. Our contribution is \textbf{methodological}: a replicable, cumulative participatory protocol that turns the abstract goal ``transparency for trust'' into prioritised, scenario-specific, and implementation-oriented requirements, together with the reasoning that makes each requirement decision-useful. We deliberately foreground the method over the numbers. A single hospital panel cannot yield population estimates, and we do not claim it does; instead we show that the protocol can produce decision-relevant evidence (such as stable priorities, a reliance-calibrated gradient, a catalogue of accepted/contested solutions) and that we argue is part of what an institution needs to know about AI transparency perception for clinical AI deployment at the benefit of the healthcare system and patient.
 
\paragraph{Contributions.}
(1) We frame clinical-AI transparency as an instrument for calibrated reliance and derive from the reliance literature why requirements must be co-defined and context-specific. (2) We present a replicable six-workshop methodology, with its instruments and analysis, designed to be transferable across institutions. (3) We report a European university-hospital use case as an existence proof, reading the results as signals about the method rather than as proven findings. (4) We discuss what the method enabled locally, and its limits, for GenAI adoption and governance.

\begin{figure}[t]
  \centering
  \includegraphics[width=\linewidth]{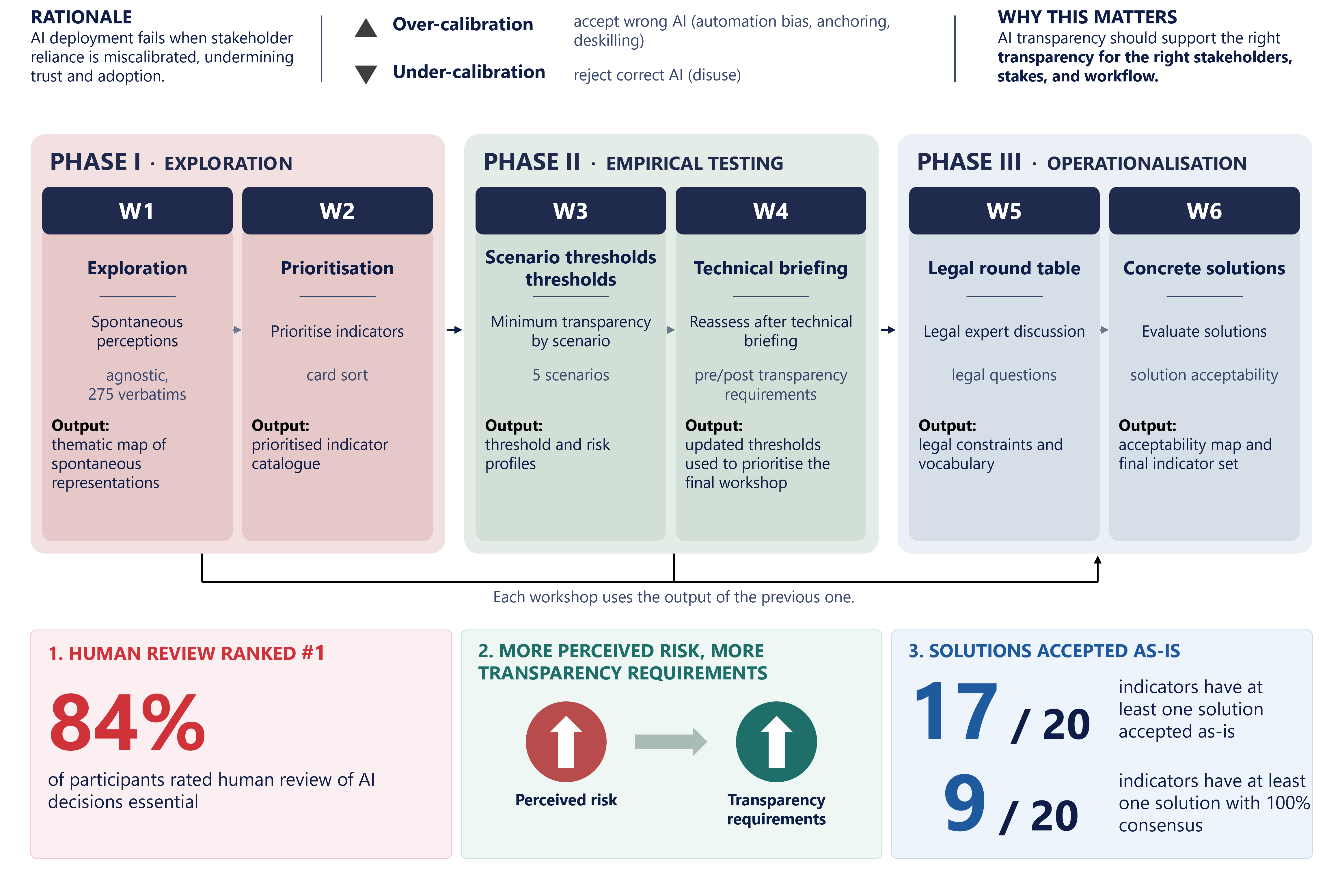}
  \caption{\textbf{The cumulative six-workshop methodology, in three phases.} Each card summarises the workshop’s purpose and output; the output of one workshop is the input of the next. The chain converts agnostic and open-ended perceptions into prioritised, scenario-specific, legally-informed and operationalisation-oriented transparency requirements.}
  \label{fig:pipeline}
\end{figure}
 
\section{Background and related work}
\label{sec:background}
 
Appropriate reliance is the human's ability to accept correct AI advice and reject incorrect advice \citep{schemmer2023,leesee2004,parasuraman1997}. It decomposes into avoiding over-reliance (following the AI when it is wrong) and avoiding under-reliance (overriding it when it is right); good outcomes require both \citep{kuper2025,calibrating2026}. In clinical settings, the balance is fragile and expertise-dependent: novices tend to over-rely because they detect errors less well, while experts may under-rely \citep{logg2019,tschandl2020,kuper2025}. A randomised vignette study found that systematically biased model predictions degraded clinicians' diagnostic accuracy, and that standard explanations did not prevent this harm \citep{jabbour2023}; an accompanying analysis framed automation bias as a first-order safety risk \citep{khera2023}. Recent evidence extends the concern to skill retention---endoscopists' unaided detection performance fell after routine exposure to AI assistance \citep{budzyn2025}---and even AI-literacy training did not eliminate automation bias in LLM-assisted diagnostic reasoning \citep{qazi2026}. Conversely, controlled studies of second-opinion systems find disuse to be the more prevalent and consequential failure, with appropriate reliance well below chance \citep{calibrating2026}. The natural hope is that explanation calibrates reliance, however explanations do not automatically yield complementary human--AI performance \citep{bansal2021}; they frequently fail to curb over-reliance because people adopt low-effort heuristics rather than engaging analytically, and cognitive-forcing designs help only at a usability cost \citep{bucinca2021}. Whether an explanation reduces over-reliance depends on a cost--benefit calculus---it helps when it makes verification cheaper than blind acceptance \citep{vasconcelos2023}.

There is a broad ``participatory turn'' in AI design \citep{delgado2023,kallina2024}, and healthcare-specific reviews document a range of empirical methods for public and patient involvement in AI research \citep{frost2024,ppie2023}, with a growing case for inclusive co-creation in digital health \citep{nickel2024}. Two constituencies are easy to omit yet central to calibrated reliance. \textbf{Patients and informal caregivers} are increasingly the direct users of generative tools (symptom chatbots, portals) and bear the consequences of clinician reliance; their notion of ``enough transparency'' rarely matches a model card. \textbf{Students and trainees} are the population whose reliance habits are being formed now and who are most vulnerable to over-reliance and deskilling \citep{budzyn2025,kuper2025}; embedding them in requirement-setting is simultaneously a data source and an AI-literacy intervention, complementing efforts to build AI literacy into medical curricula \citep{huang2025,singla2024}.

Transparency is routinely invoked as an ethical, regulatory or technical requirement, yet it is rarely defined together with the full range of people it concerns. Existing efforts are mostly top-down (expert definitions, checklists, XAI metrics) or one-off (an isolated focus group, a public consultation without follow-up), and the few participatory approaches remain tied to a single product or use case. Classical technology-acceptance models---the Technology Acceptance Model \citep{davis2024tam} and the Unified Theory of Acceptance and Use of Technology \citep{venkatesh2003utaut}---identify the levers of adoption (perceived usefulness, perceived ease of use, social influence, facilitating conditions), and their relevance to healthcare is well documented \citep{holden2010tam,kuhne2025}. They explain, however, \emph{why} a system is accepted or rejected rather than \emph{what} should change for a given audience. Value-Sensitive Design (VSD) \citep{friedman2013vsd} takes the opposite route: it starts from stakeholders' values (autonomy, intelligibility, safety, fairness, transparency), translates them into concrete design criteria (which information, in what form, at what moment), and involves users throughout the design process. Our methodology follows this logic, applied to transparency and carried across a cumulative sequence of workshops rather than a single product.

Our aim is not to replace FUTURE-AI \citep{futureai2025}, the EU AI Act \citep{euaiact2024}, the FMTI \citep{fmti2024}, national e-government standards, ISO/IEC~42001 \citep{iso42001} or WHO guidance \citep{who2021,who2023,who2024}, but to operationalise them: to determine which of their many provisions a given stakeholder community treats as essential, how that changes with clinical stake, and which concrete implementations they will accept. The presented methodology co-defines AI transparency with a broad set of profiles (clinicians, technical staff, patients, caregivers and students) at the same table,  in order to find solutions to operationalise transparency solutions and indicators in health institutions. The protocol borrows established components (open-ended focus groups and grounded-theory coding, MoSCoW card-sorting, iterative rating as in Delphi studies, and VSD), but combines them differently. Unlike a Delphi or nominal-group process, it is deliberative and mixes lay and expert participants at the same table. We chose this design because AI transparency must be acceptable to non-experts, yet feasible and lawful.

\section{Methods}
\label{sec:method}

\subsection{Design principles}
Four principles shape the protocol. \textbf{(P1) Agnostic/open-ended start.} The first workshop imposes no framework or technical primer, so that spontaneous stakeholder representations---not the facilitators' ontology---seed the indicator set. \textbf{(P2) Cumulative chain.} The six workshops are not independent modules; each consumes the previous output and feeds the next (Figure~\ref{fig:pipeline}), so priorities set early are stress-tested against scenarios, technical knowledge, law and concrete solutions. \textbf{(P3) Reliance-oriented elicitation.} Requirements are elicited against concrete uses that vary the depth of AI intervention, and are phrased as \emph{minimum transparency needed to accept the use}, aligning the measure with use-case AI transparency solutions acceptability rather than with abstract approval. \textbf{(P4) Every verdict is a result.} Whether a solution is judged satisfactory, in need of rework, or unacceptable, all three are informative for design; the goal is to document what is acceptable, not to validate a predetermined best practice.
 
\subsection{The six workshops}

In high-pressure environments such as university hospitals, getting time and longitudinal adherence from stakeholders is hard. The 6 workshops are designed to get the maximum valuable information from a minimal timeframe (such as the lunch break), while staying comfortable enough to permit reflection and conversations toward either a consensus or a clearer understanding of the divergent perceptions. 

\textbf{W1 -- Exploration (agnostic and open-ended).} Stakeholders answer seven open prompts on sticky notes (what they like/dislike about AI; candidate transparency and comprehension indicators; solutions; free remarks), followed by group discussion. A dense pre-workshop questionnaire anchors individual baselines; a post-workshop questionnaire captures cool-down reflections. Verbatims are analysed with open, axial and selective coding in the grounded-theory tradition \citep{strausscorbin1990}, with three independent coders reconciled collectively as triangulation. \emph{Output:} a thematic map of spontaneous representations for AI transparency.

\textbf{W2 -- Prioritisation (card-sort).} Candidate indicators are distilled from three external references (the Swiss e-government standard eCH-0272, the EU AI Act \citep{euaiact2024}, and the Foundation Model Transparency Index \citep{fmti2024}) cross-referenced and reduced to a working catalogue, each item tagged with its source. In a simultaneous individual card-sort, participants place each randomly-ordered indicator into a MoSCoW-style tier (\emph{Essential} / \emph{Useful but not essential} / \emph{Accessory}) \citep{clegg1994case}, with a brief argument round after each card. Two rules govern the sort: ``the impossible does not exist'' (judge relevance independently of feasibility) and ``AI in general'' (flag any indicator that participants think must be treated differently by system type). \emph{Output:} a prioritised catalogue, grouped post-hoc into \emph{interactional}, \emph{algorithmic} and \emph{social} families \citep{haresamudram2023three}.
 
\textbf{W3 -- Minimum thresholds per scenario.} Five narrative scenarios span the depth of AI intervention (Table~\ref{tab:scenarios}). For every indicator and every scenario, participants rate perceived risk (1--5) and the \emph{minimum transparency required to accept the use} (1--10); presentation order is counter-balanced across groups. \emph{Output:} per-scenario, per-indicator threshold and risk profiles; separates indicators whose importance is context-sensitive from those that are context-stable.

\textbf{W4 -- Effect of technical knowledge.} The same instrument is re-administered after a short ($\leq$20-minute) briefing on the ``black box,'' post-hoc explainability \citep{ribeiro2016why,lundberg2017unified, turbe2023evaluation}, interpretable-by-design models \citep{turbe2024protos}, retrieval-augmented generation \citep{lewis2020retrieval} and chain-of-thought \citep{wei2022chain, korbak2025monitorability}, with each method's limits. Because W4 measures thresholds after participants understand what transparency tools can and cannot deliver, its values (not W3's) are used to prioritise the final workshop. \emph{Output:} pre/post threshold change following the technical briefing.

\textbf{W5 -- Legal round table.} Rather than a focus group, domain lawyers answer stakeholder questions raised in W3--W4, organised into themes (liability and causation under algorithmic opacity; medical-device status under Medical Device Regulation (MDR) \citep{eu_mdr_2017}; the limits of legally required explainability \citep{wachter2017why}; medical vs.\ industrial secrecy; secondary use of health data and consent \citep{gdpr2016,ehds2024}; hospital governance; expected legal evolution). \emph{Output:} legal constraints and vocabulary that are then cited inside concrete solutions.
 
\textbf{W6 -- Concrete solutions.} For each prioritised indicator, the team proposes implementation solutions---some a \emph{common baseline} (identical across scenarios) and some organised as a per-scenario matrix. Participants vote each solution on a 0/1/2 scale (unacceptable / to rework / satisfactory). Participants discussed the proposals in mixed-profile subgroups, assigned a consensus rating to each solution, and could suggest modifications or alternative solutions. Indicators recovered at a later stage (data confidentiality; patient information, consent and choice) are evaluated in parallel to decide their adoption and priority. Solutions explicitly encode W5's legal lessons. \emph{Output:} an acceptability map over concrete solutions, and a final indicator set aligned to FUTURE-AI \citep{futureai2025}.
 
\begin{table}[t]
  \caption{The five scenarios span increasing depth of AI intervention in the care pathway. They are analysed separately to expose context-discriminating requirements, though in practice a single pathway may chain them.}
  \label{tab:scenarios}
  \centering
  \small
  \begin{tabular}{llp{7.4cm}}
    \toprule
    & Locus of influence & Illustrative use \\
    \midrule
    S1 & Patient's knowledge & Public symptom/care chatbot (patient or caregiver view) \\
    S2 & Clinician's knowledge & Internal document-synthesis tool consulted around a visit \\
    S3 & The \emph{when} & Triage / prioritisation of who is seen first \\
    S4 & The \emph{what} & Diagnostic proposal from collected clinical data \\
    S5 & The \emph{how} & Treatment recommendation / therapeutic follow-up \\
    \bottomrule
  \end{tabular}
\end{table}

The six-workshop structure was defined in advance, while W1 and subsequent workshop outputs progressively shaped the content and format of downstream workshops. More fundamentally, Workshops~2--6 were not fixed in advance: they were \emph{derived} from the themes W1 surfaced. Because W1 is deliberately agnostic and open-ended, it functions as a local needs-assessment whose emergent concerns dictate what the rest of the chain must do---rank them (W2), test their context-dependence (W3), probe the effect of technical understanding (W4), adjudicate their legal dimension (W5) and turn them into concrete solutions (W6). This makes the chain \emph{site-relative}: at another institution W1 may foreground different dominant concerns, and the downstream workshops should follow. A workshop may be dropped, merged or replaced (a site where algorithmic fairness or reimbursement dominates might add a dedicated audit workshop; a panel indifferent to environmental impact might retire those indicators; a jurisdiction with settled case law might fold the legal round table into a short briefing. What transfers is therefore not the particular six workshops or their numbers but the \emph{derivation rule}: run an agnostic and open-ended elicitation, then let its themes define the chain. We treat W1 as the methodological core and the process. 

\subsection{Analysis and outputs}
The protocol yields both quantitative and qualitative streams: card-sort tier proportions (W2); per-scenario risk and threshold means with non-parametric context tests (W3--W4); pre/post threshold deltas (W4); solution-acceptability distributions (W6); and a pre/post participant questionnaire matched by individual identifiers for longitudinal comparison. Each numeric result is paired with a discussion explaining \emph{why} participants set a threshold or rejected a solution, so that the outputs are actionable rather than merely descriptive.

The core methodological principles are not site-specific. The external references at W2 can be swapped for the jurisdiction of interest; the scenarios are templates parameterised by care pathway; the legal round table adapts to local law; and the solution catalogue is meant to be re-authored per institution. The method is specified as a step-by-step facilitator guide with fixed timings, materials and analysis recipes, so a different hospital or a non-clinical GenAI setting can run it with fidelity while adapting content.

\section{Use case: a European university hospital}
\label{sec:usecase}

 \begin{figure}[t]
  \centering
  \includegraphics[width=\linewidth]{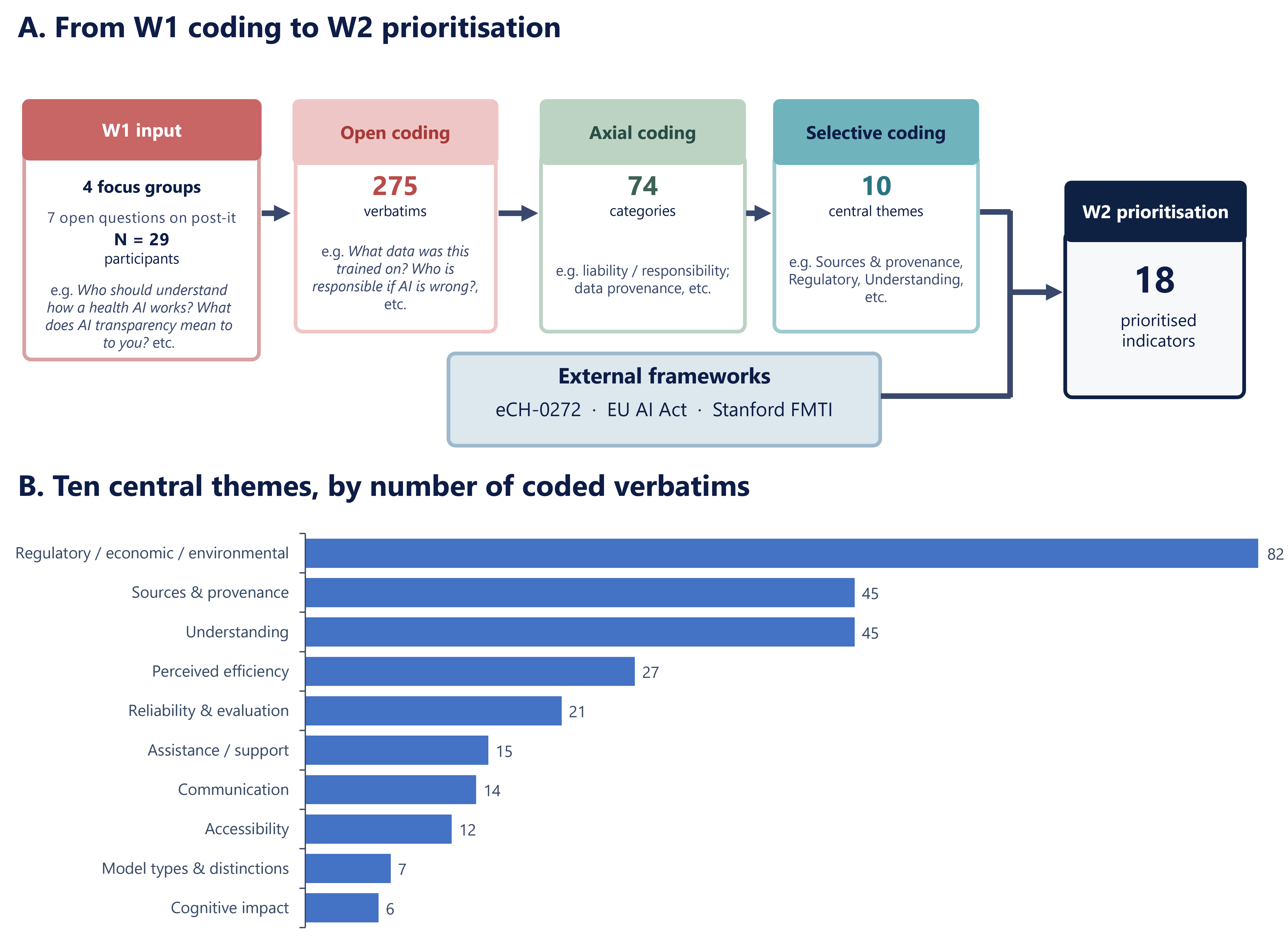}
  \caption{\textbf{W1 coding and transition to W2.} \emph{A:} The grounded-theory pipeline from four focus groups and seven open prompts, through independent open coding of the verbatims, axial grouping into categories, and selective coding reconciled across three coders, feeding the cross-referencing at W2 that yields the 18 indicators. \emph{B:} The ten central themes by number of coded verbatims; regulatory/economic/environmental concerns, sources and understanding dominate, with a marginal tail.}
  \label{fig:w1coding}
\end{figure}

We instantiated the methodology at Geneva University Hospitals. We stress the framing of Section~\ref{sec:method}: with a single self-selected panel these results are an \emph{existence proof} that the method produces decision-useful evidence, not population estimates. We therefore report effects qualitatively where the sample is small and flag uncertainty explicitly.
 
\paragraph{Panel.} Of 44 applicants, 31 were retained; to preserve diversity after attrition, four participants joined from W4 onward, and attendance ranged 20--29 per session. The panel spanned 11 health professionals, 4 software developers, 2 communication and 4 administrative staff, 1 researcher, 4 patient partners, 1 informal-caregiver partner, 1 parent partner and 3 medical students, an inter-professional mix by design (\S\ref{sec:background}). At each workshop, participants were divided into
two to four mixed-profile focus groups of roughly five to ten people, each session
lasting about one hour. Profiles were mixed \emph{within} groups rather than
segregated, so that every subgroup carried several viewpoints; presentation
order of materials was counter-balanced across groups where relevant. In the
solution-evaluation workshop (W6), each of the four groups was further split
into two subgroups, yielding eight subgroups whose votes were pooled.
 
\paragraph{Baseline (W1).} All respondents (29/29) had used AI personally and 24/29 professionally, yet 13/29 (45\%) had never encountered the notion of AI \emph{transparency}; nonetheless 25/29 rated it important or very important. Coding of 275 verbatims produced 74 categories under 10 themes; the most cited spontaneous indicators were source provenance, disclosure of AI use, the nature of data used, and the existence of a human review (Figure~\ref{fig:w1coding}).

\begin{figure}[t]
  \centering
  \includegraphics[width=0.99\linewidth]{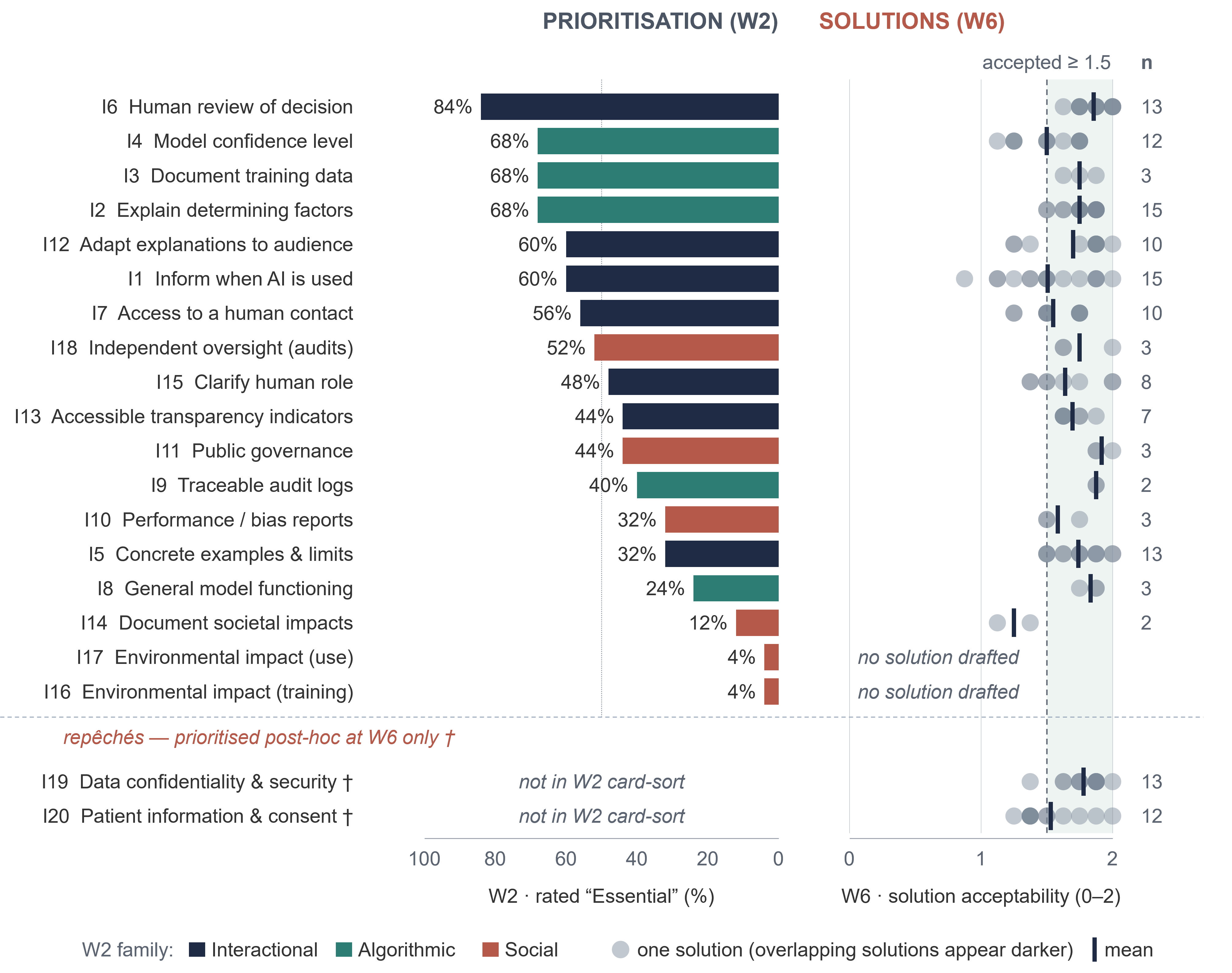}
  \caption{From prioritisation (W2) to evaluated solutions (W6), one row per indicator. \emph{Left:} share rating the indicator ``Essential'' in the card-sort, coloured by family (interactional/algorithmic/social). \emph{Right:} mean acceptability of that indicator's concrete solutions (0 = not acceptable, 1 = to rework, 2 = satisfactory), dark tick is indicator mean; shaded zone is considered as an acceptable solution as-is (score above 1.5), with $n$ the number of solutions evaluated ($n=147$ in total; 4 participant-proposed solutions excluded from figure). Grey dot is one solution's mean acceptability across 8 subgroups . Below the dashed rule, the two \emph{repêchés} indicators (I19 data confidentiality, I20 patient consent; marked $\dagger$) were introduced and prioritised only at W6 and have no W2 card-sort score.}
  \label{fig:cardsort}
\end{figure}

\paragraph{What stakeholders prioritise (W2).} From 26 candidates, 18 indicators were retained. The card-sort (Figure~\ref{fig:cardsort}) shows a clear hierarchy: \emph{the ability to request human review of an automated decision} is essential far above all else (84\% Essential, 0\% Accessory), followed by explaining the determining factors, documenting training data and stating model confidence (68\% each). Environmental-impact indicators sit at the bottom (4\% Essential). The ordering places human-in-the-loop and intelligibility (the levers most tied to reliance) above institutional and technical disclosures.
 
\begin{figure}[t]
  \centering
  \includegraphics[width=0.99\linewidth]{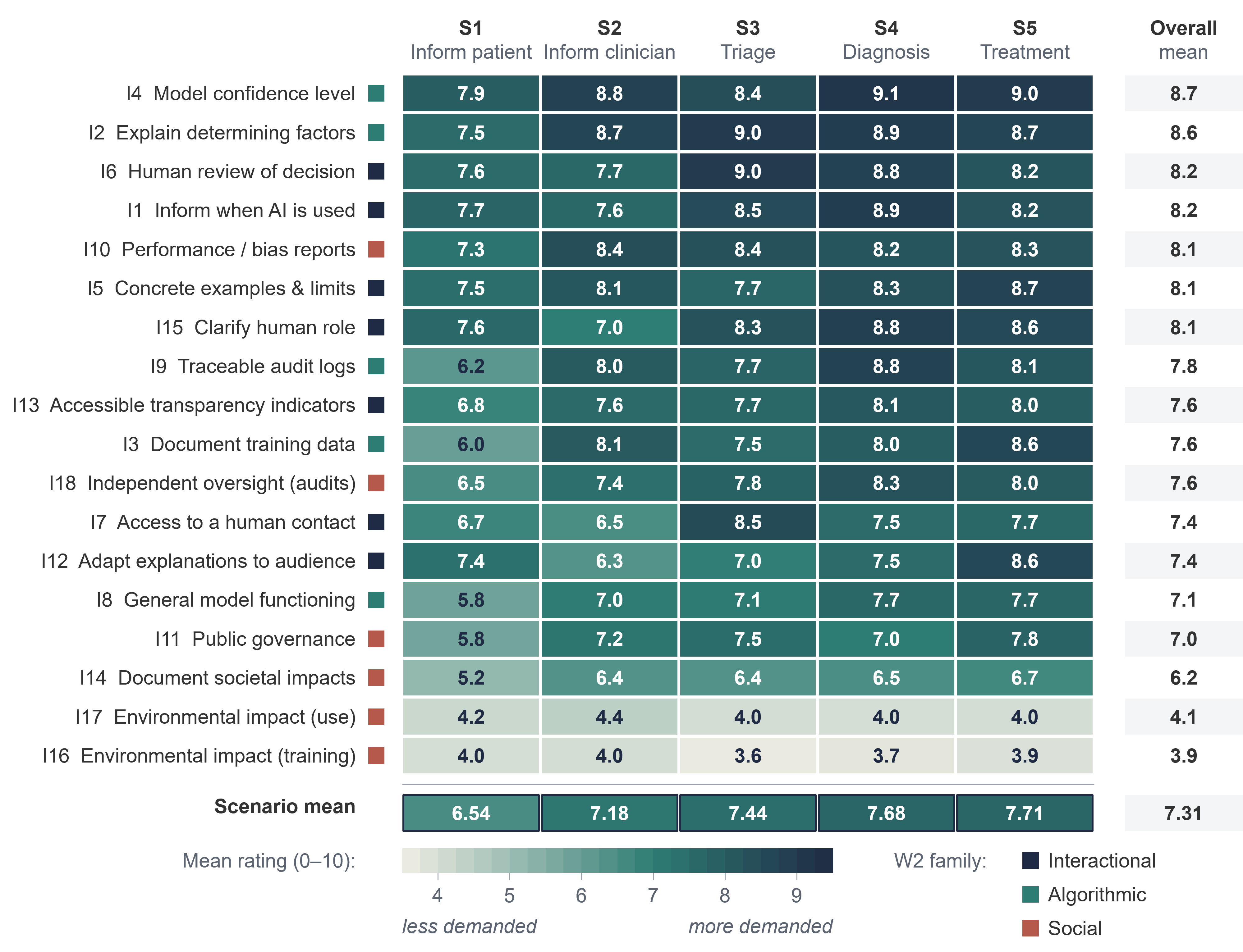}
  \caption{Minimum transparency required to accept each use after the technical briefing (W4, 1–10 scale), per indicator and scenario, with the per-indicator mean, 27 participants. Squares indicate the W2 family (interactional, algorithmic, social).}
  \label{fig:gradient}
\end{figure}

\paragraph{A reliance-calibrated gradient (W3--W4).} Required minimum transparency for adoption rises monotonically with the depth of AI intervention (from informing a patient, to informing a clinician, to triage, diagnosis and treatment; see Figure~\ref{fig:gradient}) mirroring a proportionality-to-risk logic \citep{who2023}. Perceived risk peaks earlier, at triage, while the transparency requirement keeps climbing toward treatment: the two are related but not identical, and transparency tracks \emph{consequence for the patient} more than momentary risk. The scenarios also surfaced qualitatively distinct concerns that a scalar score hides: anticipated deskilling for the clinician-facing tool (S2) \citep{budzyn2025}, anchoring once a diagnosis is proposed (S4) \citep{jabbour2023}, and the shift of human oversight from a safety net to a precondition in triage (S3).
 
\paragraph{Technical literacy does not reassure (W4).} Re-rating the same scenarios after a short ($\leq$20-minute) briefing on what explainability methods can and cannot do produced no detectable quantitative change: per-scenario shifts were small (about $+0.1$ to $+0.2$, at most $+0.25$ on the 10-point scale) and far below the response dispersion (per-indicator SD $\approx$2--4), so we do not read them as a genuine change in thresholds (Figure~\ref{fig:gradient} and Table~\ref{tab:w3w4}). The informative result was qualitative: understanding the possibilities \emph{and limits} of explainability did not relieve participants' concerns, and it did not meaningfully change what they asked for. For this audience, transparency needs appear anchored in the clinical stake and the demand for human oversight rather than in technical familiarity. This argues against treating literacy training as a substitute for building transparency into the system, and is consistent with automation bias persisting under AI literacy \citep{qazi2026}.
 
\paragraph{From principles to solutions (W6).} Of 151 solution evaluations, 79 were accepted as-is, 44 flagged for discussion, 26 for rework and 2 rejected; 14 reached perfect unanimity. Rejections and rework are as informative as approvals (P4): e.g.\ a proposal to record AI involvement in the prescription/care plan was rejected, and a right-to-a-human-contact solution weakened precisely at the highest-stake scenarios where participants doubted its feasibility. The two recovered indicators behaved sensibly: data confidentiality escalated to top priority in the more engaging clinical scenarios, whereas patient-consent priority was context-dependent. Several accepted solutions carried the legal round table's fingerprints (data-handling notices, an AI-governance charter, reinforced external audit). Reading the two workshops together (Figure~\ref{fig:cardsort}) shows the principle-to-solution link is real but not mechanical: nearly every prioritised indicator also yielded solutions judged acceptable (mean vote $\geq$1.5 on the 0--2 scale), and human review---top-ranked in W2---was also among the most acceptable. Yet priority does not dictate acceptability: a low-priority indicator (general model functioning) drew highly acceptable solutions, while the one indicator whose solutions still need rework (documenting societal impacts) was itself low-priority. Environmental indicators, deprioritised throughout, attracted no drafted solution at all. This absence is itself a finding.
 
\paragraph{Longitudinal trust.} We compare two cross-sectional
snapshots: a pre-workshop baseline (T1, $N=29$) and a post-workshop
questionnaire (T2, $N=22$) (not a matched within-participant sample), so these are descriptive signals rather than a paired change (collection incomplete at analysis). Trust in clinical AI stayed moderate at both times: the two middle categories (``moderately'' and ``fairly confident'') dominated and grew (72.4\% at T1 $\rightarrow$ 90.9\% at T2), while the extreme categories emptied and the share reporting \emph{low trust} was smaller at T2 (20.7\% $\rightarrow$ 9.1\%). We read this less as a rise in trust than as a shift toward a more conditional, reasoned trust, which is consistent with the verbatims, where T1 responses stress knowing the sources and the fact that AI is used, whereas T2 responses invoke traceability, auditability, human oversight and system documentation.

\section{Discussion}
\label{sec:discussion}
 
\paragraph{What the method enabled} The use case shows the protocol delivering four things a checklist cannot: (i) a \emph{ranked} set of requirements anchored in what stakeholders treat as essential; (ii) a \emph{context function}---how the requirement scales with clinical stake---rather than a single threshold; (iii) a pre/post assessment of whether \emph{technical information} changes requirements, useful for deciding whether to invest in literacy vs.\ built-in transparency; and (iv) an \emph{acceptability map} over concrete implementations, including the instructive failures. Each output maps directly onto a deployment decision (what to build first, where to demand more, what users will reject), which is what the GenAI4Health adoption agenda needs and what generic transparency guidance leaves under-specified.
 
\paragraph{Relation to the reliance literature.} The participants' stated priorities are consistent with findings from the reliance literature: the levers they rank highest (human review, intelligible factors, confidence) are those tied to \emph{when to trust}, and their appetite for transparency grew rather than shrank with technical understanding---echoing that explanation is not a reliability guarantee \citep{bansal2021,bucinca2021,jabbour2023}. Involving students and patients was not tokenistic: trainees' deskilling fears (S2) and patients' consent boundaries (S1/S4) shaped the scenario-specific requirements, supporting the case for co-definition (\S\ref{sec:background}).

\section{Limitations}
\emph{Statistical power}: a single site with 20--29 participants per session cannot support inferential generalisation; scenario means and deltas are indicative, and small differences should not be over-read.
\emph{Technical literacy}: was probed with a single and very short 20-minute briefing; a longer or differently delivered intervention might shift requirements, so the null result concerns this dose and format only, and should be studied more specifically.
\\ \emph{Self-selection}: volunteers are likely curious about AI.\\
\emph{Single institution and jurisdiction}: priorities and legal framing are country-specific; the \emph{method} transfers, the \emph{local outcomes} may not. \\
\emph{Declared thresholds}: workshop-elicited minimums are stated preferences, not measured reliance behaviour; connecting them to behavioural over-/under-reliance is future work.  \\
\emph{Longitudinal data} is incomplete $N=22$ out of $N=29$. \\
These limits are the reason we present the contribution as a methodology validated by demonstration rather than as a powered study. 

\paragraph{Future work:} within a single institution, the next step closes the loop from stated to behavioural reliance: instrument AI tools already deployed at the hospital with the accepted solutions, measure appropriate-reliance outcomes \citep{schemmer2023,calibrating2026}, and scale elicitation through an institution-wide survey seeded by the workshop catalogue. Across institutions, running the same protocol at multiple sites would show which indicators and perceptions are stable and which vary. Because clinical AI is deployed globally, understanding these regional dependencies (such as culture, legal frameworks, and care local processes) is key to designing transparency that is itself calibrated to the setting, and thus to calibrated reliance on AI tools.

\section{Conclusion}

Transparency keeps a central role in trustworthy clinical AI in calibrating
reliance, trust and adoption. What counts as the right transparency depends on who uses the system, how much is at stake, and where it sits in the workflow. Rather than settle this with a fixed checklist, we built a participatory methodology that elicits the requirements from the
people who act on the system's output. In our instantiation, 275 agnostic and open-ended stakeholder
verbatims were read against three external references (the Swiss e-government standard eCH-0272, the EU AI Act, and the Stanford Foundation Model Transparency
Index) to assemble a working set of transparency indicators. Participants then rated those
indicators across five clinical scenarios of increasing algorithmic involvement, before and
after a short briefing on what explainability tools can and cannot do. A legal round table
tested the requirements against liability, consent and governance, and a final workshop
turned the surviving priorities into concrete, votable implementation features. And indicators about environmental cost were consistently set aside next to immediate clinical stakes.

The single-site study produced evidence an institution can act on, and two needs stood
out. Human review emerged as the highest-priority safeguard in W2 (84\% of participants rated it Essential) and it remained highly rated across all five clinical scenarios. The second is proportionality: the
minimum transparency people asked for rose steadily with the depth of the intervention,
from roughly $6.5/10$ for a public information tool to above $7.7/10$ for
treatment-influencing AI, in line with WHO guidance that scrutiny should follow risk
\citep{who2023}. Two secondary results are worth keeping in view. Teaching participants
how explainability methods work neither reassured participants nor changed their requirements, so
technical literacy seems not to be a substitute for building transparency into the system. Of the 151 concrete solutions evaluated, 79 were accepted without change, and the exercise closed with a catalogue of 20 indicators mapped onto the FUTURE-AI framework
\citep{futureai2025}.

We therefore offer the six-workshop protocol, and not the particular figures from one
hospital, as the part worth reusing. The numbers are local; the way of producing them might be implemented in another healthcare facility. To make that usable, we release the full instruments, the coding scheme and the analysis recipes, so other teams can run the same elicitation in their own setting, adapt the scenarios and legal questions to their jurisdiction, and reach requirements their own stakeholders will actually accept.

\begin{ack}
We are grateful to all participants (health professionals, patient, caregiver and parent partners, students and staff) for their time and thoughtful contributions. We thank medical student Marius Bercher for his contribution to the organisation and analysis of the sixth workshop (W6). The XpliCAI project is funded by the Private Foundation of the Geneva University Hospitals (HUG).
\end{ack}

\bibliographystyle{plainnat}
\bibliography{references}

\newpage

\appendix
\section{Supplementary Details}

\noindent The primary contribution of this paper is a \emph{transferable method}. To let other
teams replicate it quickly and enable multi-site validation, we release the complete
facilitator guide, every instrument, the coding scheme, and the anonymised result
workbooks. This appendix summarises those materials and reports the full per-indicator data
behind the figures. The paper is self-contained; the appendix is optional reading.

% ----------------------------------------------------------------------------
\section{Released replication package}
\label{app:package}
All materials to replicate the study are provided as supplementary files (a single archive; code and data as
plain \texttt{.docx}/\texttt{.pptx}/\texttt{.xlsx}/\texttt{.csv}) at \url{https://doi.org/10.5281/zenodo.23652465}. They are language- and
site-agnostic templates: a replicating team edits the content (indicator wording, scenarios,
legal questions, solutions) while keeping the structure fixed.

\begin{table}[h]
\centering\footnotesize
\begin{tabular}{@{}p{0.30\linewidth}p{0.62\linewidth}@{}}
\toprule
\textbf{Material} & \textbf{Purpose} \\
\midrule
Methodological guide & Master facilitator manual: objectives, timings, materials, data collected, recommended analysis and deliverables for all six workshops. \\
W1 pre-workshop survey & Individual baseline: prior AI exposure, self-rated understanding, spontaneous word associations, importance of transparency, risks of opaque AI. \\
W1 post-workshop survey & Cool-down reflections, matched to the pre-survey by pseudonymous ID. \\
W2 indicator deck & The candidate indicators used in the card-sort, each tagged with its source framework. \\
W3 evaluation questionnaire & Per-indicator, per-scenario rating sheet (risk 1--5; minimum transparency 1--10). \\
W3/W4 scenario booklet & The five narrative scenarios of graduated AI intervention (Table~\ref{tab:scenarios} in the main text). \\
W4 training deck & The $\leq$20-minute ``black-box'' briefing (post-hoc explainability, interpretable-by-design, retrieval-augmentation, chain-of-thought, and each method's limits). \\
W5 legal-expert brief & Instructions and the question set for the domain-lawyer round table. \\
W6 solution survey \& post-survey & Solution catalogue with the 0/1/2 acceptability ballot, plus the final questionnaire. \\
Result workbooks (anonymised) & W1 coding crosswalk (verbatim $\rightarrow$ category $\rightarrow$ theme), W2 card-sort proportions, W3/W4 threshold and risk means, and the W6 solution-level votes with recomputed status. \\
\bottomrule
\end{tabular}
\caption{Contents of the released replication package.}
\label{tab:package}
\end{table}

% ----------------------------------------------------------------------------
\section{General format, panel and ethics}
\label{app:format}
\textbf{Panel.} We recommend roughly 30 participants, split into three to four groups of
five to ten, chosen for stakeholder diversity (health professionals, technical staff,
administrative/communication staff, patient and caregiver partners, and students/trainees)
rather than statistical representativeness. Group size can be adapted to local resources; the
invariant is diversity plus small-enough groups for active participation. Each workshop runs
60 minutes.

\textbf{Ethics, consent and data management.} Ethical and institutional requirements are
verified locally before recruitment, and consent follows the procedures of the host
institution. Each participant receives a pseudonymous identifier (\texttt{p01}, \texttt{p02},
\dots); the table linking identities to identifiers is stored separately from research data
and accessible only to authorised staff. The same identifier matches pre/post questionnaires
without revealing identities in the analysed data.

\textbf{Fidelity vs.\ adaptation.} Replicating teams are asked to keep the \emph{sequence},
the \emph{elicitation logic} (agnostic and open-ended start; minimum-transparency-to-accept framing;
before/after technical briefing; ``every verdict is a result'') and the analysis recipes,
while adapting the \emph{content}: indicator wording and terminology, the clinical scenarios,
the legal material, and the implementation solutions.

% ----------------------------------------------------------------------------
\section{Facilitator guide (summary)}
\label{app:facilitator}
Table~\ref{tab:facilitator} condenses the per-workshop protocol; full scripts, slides and
sheets are in the released package (\S\ref{app:package}).

\begin{table}[h]
\centering\footnotesize
\begin{tabular}{@{}p{0.055\linewidth}p{0.055\linewidth}p{0.45\linewidth}p{0.34\linewidth}@{}}
\toprule
\textbf{WS} & \textbf{Time} & \textbf{Phases} & \textbf{Output} \\
\midrule
W1 & 60\,min & Pre-survey (10--15'); self-introductions; individual reflection on seven posted prompts, one idea per sticky note written \emph{before} discussion (10--15'); group discussion (20--25'); post-survey emailed. & Thematic map of spontaneous representations (open/axial/selective coding, three coders). \\
W2 & 60\,min & Intro (5'); rating categories (5'); individual card-sort of each indicator into Essential / Useful / Accessory with an argument round (45'); post-hoc grouping into interactional/algorithmic/social families; summary (5'). & Prioritised indicator catalogue. \\
W3 & 60\,min & Intro (5'); for each of five scenarios, rate perceived risk (1--5) and minimum transparency to accept (1--10) per indicator, order counter-balanced (50'); conclusion (5'). & Per-scenario, per-indicator threshold and risk profiles. \\
W4 & 60\,min & Intro (5'); technical briefing on explainability and its limits (20'); re-rate the same scenarios (25'); group discussion (10'). & Pre/post changes in transparency requirements following a technical briefing. \\
W5 & 60\,min & Intro (5'); summary of the legal issues raised in W3--W4 (10'); moderated discussion with domain lawyers (45'). & Legal constraints and vocabulary, cited inside W6 solutions. \\
W6 & 60\,min & Present a common baseline and per-scenario solutions per prioritised indicator; vote each on 0/1/2 (unacceptable / to rework / satisfactory); evaluate two recovered indicators; finalise. & Acceptability map over solutions; final indicator set. \\
\bottomrule
\end{tabular}
\caption{Six-workshop facilitator guide (each session 60 minutes).}
\label{tab:facilitator}
\end{table}

% ----------------------------------------------------------------------------
\section{Workshop 1 coding}
\label{app:w1coding}
Verbatims were analysed with open, axial and selective coding (grounded theory), three
coders reconciled by consensus. Coding of 275 verbatims produced 74 categories under the ten
central themes in Table~\ref{tab:themes}.

\begin{table}[h]
\centering\footnotesize
\begin{tabular}{@{}p{0.34\linewidth}r p{0.50\linewidth}@{}}
\toprule
\textbf{Theme} & \textbf{Coded} & \textbf{What it captures} \\
\midrule
Regulatory / economic / environmental & 82 & governance, data misuse, sustainability, job impact \\
Sources \& provenance & 45 & where answers come from; verifiability \\
Understanding & 45 & how the AI works and its limits \\
Perceived efficiency & 27 & speed, usefulness, information overload \\
Reliability \& evaluation & 21 & performance, bias, trustworthiness \\
Assistance / support & 15 & decision support, not replacement \\
Communication & 14 & disclosing that AI is used; clarity \\
Accessibility & 12 & plain language, adapting to audiences \\
Model types \& distinctions & 7 & different AI kinds behave differently \\
Cognitive impact & 6 & deskilling, loss of critical judgement \\
\bottomrule
\end{tabular}
\caption{Workshop~1 central themes, by number of coded verbatims.}
\label{tab:themes}
\end{table}

% ----------------------------------------------------------------------------
\section{Indicator catalogue}
\label{app:catalogue}
The card-sort retained 18 indicators from 26 candidates; two further indicators (I19, I20)
were reintroduced at W6 because they recurred throughout but were not covered by the initial
set, giving 20 in the final catalogue. Table~\ref{tab:catalogue} lists them with their W2
priority (\% rating ``Essential''), their mean W4 threshold (averaged over the five
scenarios, 1--10), and their W6 solution acceptability (mean of 0--2 votes; number of
solutions in parentheses). Family: Int(eractional), Alg(orithmic), Soc(ial). A detailed
cross-walk of every indicator to the external frameworks (the Swiss e-government standard eCH-0272, the EU AI Act, the Stanford Foundation Model Transparency Index,
FUTURE-AI and ISO/IEC~42001) is provided in the result workbook; the two environmental
indicators (I16/I17) have no dedicated FUTURE-AI counterpart and are an original contribution
of this catalogue.

\begin{table}[h]
\centering\footnotesize
\begin{tabular}{@{}llcccc@{}}
\toprule
\textbf{ID} & \textbf{Indicator} & \textbf{Fam.} & \textbf{W2 \%Ess} & \textbf{W4 mean} & \textbf{W6 acc. ($n$)} \\
\midrule
I6 & Human review of decision & Int & 84 & 8.24 & 1.86 (13) \\
I4 & Model confidence level & Alg & 68 & 8.66 & 1.50 (12) \\
I2 & Explain determining factors & Alg & 68 & 8.56 & 1.75 (15) \\
I3 & Document training data & Alg & 68 & 7.63 & 1.75 (3) \\
I1 & Inform when AI is used & Int & 60 & 8.20 & 1.51 (15) \\
I12 & Adapt explanations to audience & Int & 60 & 7.35 & 1.70 (10) \\
I7 & Access to a human contact & Int & 56 & 7.40 & 1.55 (10) \\
I18 & Independent oversight (audits) & Soc & 52 & 7.61 & 1.75 (3) \\
I15 & Clarify human role & Int & 48 & 8.05 & 1.64 (8) \\
I11 & Public governance & Soc & 44 & 7.04 & 1.92 (3) \\
I13 & Accessible transparency indicators & Int & 44 & 7.63 & 1.70 (7) \\
I9 & Traceable audit logs & Alg & 40 & 7.77 & 1.88 (2) \\
I5 & Concrete examples \& limits & Int & 32 & 8.07 & 1.74 (13) \\
I10 & Performance / bias reports & Soc & 32 & 8.10 & 1.58 (3) \\
I8 & General model functioning & Alg & 24 & 7.06 & 1.83 (3) \\
I14 & Document societal impacts & Soc & 12 & 6.21 & 1.25 (2) \\
I16 & Environmental impact (training) & Soc & 4 & 3.86 & no solution \\
I17 & Environmental impact (use) & Soc & 4 & 4.14 & no solution \\
\midrule
I19 & Data confidentiality \& security$^\dagger$ & -- & -- & -- & 1.78 (13) \\
I20 & Patient information \& consent$^\dagger$ & -- & -- & -- & 1.53 (12) \\
\bottomrule
\end{tabular}
\caption{The 20-indicator catalogue (sorted by W2 priority). $^\dagger$ recovered for
W6; not in the W2 card-sort. Single-site values.}
\label{tab:catalogue}
\end{table}

% ----------------------------------------------------------------------------
\section{Per-indicator thresholds by scenario (W3/W4)}
\label{app:w3w4}
Table~\ref{tab:w3w4} gives the minimum transparency required after the technical briefing
(W4) for every indicator in each scenario (S1 inform patient, S2 inform clinician, S3 triage,
S4 diagnosis, S5 treatment). Each cell reports the mean with its standard deviation in
parentheses; the last two columns give the per-indicator mean over scenarios and the mean
shift from W3 ($\Delta$). The scenario-level W4 means are 6.54, 7.18, 7.44, 7.68 and 7.71,
reproducing the monotonic gradient; the large SDs (typically 2--4 points on the 1--10 scale)
reflect a small, heterogeneous panel and are the reason we read these values as signals rather
than estimates. The full per-scenario SD tables for both W3 and W4 are in the released
workbook (the \texttt{Synthese\_SD} and \texttt{Synthese\_SD\_A4} sheets).

\begin{table}[h]
\centering\scriptsize\setlength{\tabcolsep}{4pt}
\begin{tabular}{@{}lccccccr@{}}
\toprule
\textbf{ID} & \textbf{S1} & \textbf{S2} & \textbf{S3} & \textbf{S4} & \textbf{S5} & \textbf{mean} & \textbf{$\Delta$} \\
\midrule
I1 & 7.7\,(3.1) & 7.6\,(3.2) & 8.5\,(2.5) & 8.9\,(2.1) & 8.2\,(2.6) & 8.20 & +0.12 \\
I2 & 7.5\,(2.9) & 8.7\,(1.5) & 9.0\,(1.5) & 8.9\,(1.4) & 8.7\,(1.8) & 8.56 & +0.04 \\
I3 & 6.0\,(2.8) & 8.1\,(2.6) & 7.5\,(3.2) & 8.0\,(2.9) & 8.6\,(1.6) & 7.63 & +0.17 \\
I4 & 7.9\,(2.5) & 8.8\,(2.3) & 8.4\,(2.4) & 9.1\,(2.2) & 9.0\,(1.6) & 8.66 & +0.11 \\
I5 & 7.5\,(2.7) & 8.1\,(1.8) & 7.7\,(2.0) & 8.3\,(2.2) & 8.7\,(1.3) & 8.07 & +0.64 \\
I6 & 7.6\,(3.1) & 7.7\,(2.8) & 9.0\,(1.2) & 8.8\,(1.9) & 8.2\,(2.4) & 8.24 & +0.21 \\
I7 & 6.7\,(3.7) & 6.5\,(3.8) & 8.5\,(2.1) & 7.5\,(3.4) & 7.7\,(3.1) & 7.40 & +0.03 \\
I8 & 5.8\,(3.2) & 7.0\,(2.8) & 7.1\,(3.2) & 7.7\,(2.6) & 7.7\,(2.6) & 7.06 & +0.27 \\
I9 & 6.2\,(3.0) & 8.0\,(1.7) & 7.7\,(2.8) & 8.8\,(1.9) & 8.1\,(2.8) & 7.77 & +0.41 \\
I10 & 7.3\,(2.3) & 8.4\,(2.3) & 8.4\,(2.0) & 8.2\,(2.0) & 8.3\,(2.1) & 8.10 & +0.44 \\
I11 & 5.8\,(3.1) & 7.2\,(2.4) & 7.5\,(2.0) & 7.0\,(2.5) & 7.8\,(1.7) & 7.04 & +0.09 \\
I12 & 7.4\,(2.7) & 6.3\,(3.5) & 7.0\,(4.0) & 7.5\,(2.9) & 8.6\,(2.3) & 7.35 & +0.01 \\
I13 & 6.8\,(3.0) & 7.6\,(1.9) & 7.7\,(1.9) & 8.1\,(1.5) & 8.0\,(2.3) & 7.63 & +0.44 \\
I14 & 5.2\,(3.4) & 6.4\,(2.9) & 6.4\,(3.7) & 6.5\,(3.4) & 6.7\,(3.3) & 6.21 & +0.03 \\
I15 & 7.6\,(2.0) & 7.0\,(2.9) & 8.3\,(1.7) & 8.8\,(0.9) & 8.6\,(1.8) & 8.05 & +0.17 \\
I16 & 4.0\,(2.8) & 4.0\,(2.9) & 3.6\,(3.1) & 3.7\,(3.1) & 3.9\,(3.0) & 3.86 & $-0.07$ \\
I17 & 4.2\,(2.7) & 4.4\,(2.9) & 4.0\,(3.0) & 4.0\,(2.9) & 4.0\,(2.8) & 4.14 & $-0.07$ \\
I18 & 6.5\,(3.2) & 7.4\,(2.6) & 7.8\,(2.6) & 8.3\,(2.7) & 8.0\,(3.0) & 7.61 & +0.13 \\
\midrule
\textbf{mean} & 6.54 & 7.18 & 7.44 & 7.68 & 7.71 & & \\
\bottomrule
\end{tabular}
\caption{Minimum transparency required per indicator $\times$ scenario after the briefing
(W4, 1--10), shown as mean\,(SD). Last columns: per-indicator mean over scenarios and mean
shift from W3 ($\Delta$). Single-site values; SDs are large, so read as signals not estimates.}
\label{tab:w3w4}
\end{table}

% ----------------------------------------------------------------------------
\section{Workshop 6 solution acceptability}
\label{app:w6}
Of 151 solution evaluations, 79 were accepted as-is, 44 flagged for discussion, 26 for rework
and 2 rejected; 14 reached perfect unanimity. Per-indicator acceptability is reported in the
last column of Table~\ref{tab:catalogue}; the solution-level ballots, standard deviations and
a rule-based status (Accepted if mean $\geq$1.5 with strong consensus; To rework if
1.0--1.5; Non-accepted if $<$1.0; To discuss if mean $\geq$1.5 but dispersed) are in the
result workbook. Rejections and rework are treated as findings: for example a proposal to log
AI involvement in the prescription/care plan was rejected, and a right-to-human-contact
solution weakened at the highest-stake scenarios where participants doubted its feasibility.

% ----------------------------------------------------------------------------
\section{Longitudinal questionnaire and statistical notes}
\label{app:stats}

\paragraph{Longitudinal (provisional).} Trust was measured at two time points
on independent cross-sections (T1 pre-workshop, $N=29$; T2 post-workshop,
$N=22$); we therefore report distributions per time point rather than a matched
delta (Table~\ref{tab:trust}). Moderate categories predominated throughout and
the low-trust share was lower at T2; qualitatively, criteria shifted from an
informational view (sources, disclosure of AI use) toward a systemic one
(traceability, auditability, human supervision, documentation).

\begin{table}[h]
\centering
\caption{Self-reported trust in the clinical use of AI, per time point
(independent cross-sections).}
\label{tab:trust}
\begin{tabular}{lcc}
\toprule
Trust level & T1 pre ($N=29$) & T2 post ($N=22$) \\
\midrule
No trust at all  & 1 (3.4\%)   & 0 (0\%) \\
Low trust        & 6 (20.7\%)  & 2 (9.1\%) \\
Moderate trust   & 11 (37.9\%) & 10 (45.5\%) \\
Fairly confident & 10 (34.5\%) & 10 (45.5\%) \\
Full trust       & 1 (3.4\%)   & 0 (0\%) \\
\bottomrule
\end{tabular}
\end{table}

\textbf{Analysis.} W2 is summarised by tier proportions; W3/W4 by per-scenario and
per-indicator means to separate context-sensitive from context-stable indicators); W4 by matched
pre/post deltas. Given a single self-selected panel of 20--29 per session, all quantities are
descriptive: they are read as signals about the method, not as population estimates.

% ----------------------------------------------------------------------------
\section{Guidance for multi-site replication}
\label{app:multisite}
For a comparable second site we suggest: (1) keep the pseudonymous-ID and pre/post-matching
scheme so results can be pooled; (2) reuse the indicator catalogue (Table~\ref{tab:catalogue})
as the W2 starting set, swapping the national standard for the local one and letting W1 add
site-specific candidates; (3) keep the five-scenario spine but re-word illustrative uses to
local care pathways; (4) reuse the risk (1--5) and transparency (1--10) scales verbatim so
gradients are comparable across sites; (5) localise the W5 legal questions to the relevant
jurisdiction. Because the chain is derived from W1, a site may add, merge or drop a workshop
(e.g.\ a dedicated fairness-audit session, or folding the legal round table into a briefing
where case law is settled). Reporting the per-indicator W2/W4/W6 values in the format of
Tables~\ref{tab:catalogue}--\ref{tab:w3w4} would make cross-site comparison and eventual
meta-analysis straightforward.

\addcontentsline{toc}{section}{Appendix}

\end{document}